\RequirePackage{fix-cm}
\documentclass[twocolumn,epjc3]{svjour3}          
\smartqed  
\usepackage{graphicx}
\usepackage{booktabs}
\usepackage{siunitx}
\usepackage{amsmath}
\usepackage{amssymb}
\usepackage[switch,mathlines]{}  
\usepackage{xspace}
\usepackage[T5,T1]{fontenc}
\usepackage[compact]{titlesec}
\usepackage[shortcuts]{extdash}

\usepackage{newtxtext,newtxmath}   

\journalname{Eur. Phys. J. C}

\begin{document}

\title{Constraining the Prompt Atmospheric Neutrino Flux Combining IceCube's Cascade and Track Samples
}

\titlerunning{Constraining the Prompt Atmospheric Neutrino Flux Using IceCube}        

\onecolumn
\author{R. Abbasi\thanksref{loyola}
\and M. Ackermann\thanksref{zeuthen}
\and J. Adams\thanksref{christchurch}
\and S. K. Agarwalla\thanksref{madisonpac,a}
\and J. A. Aguilar\thanksref{brusselslibre}
\and M. Ahlers\thanksref{copenhagen}
\and J.M. Alameddine\thanksref{dortmund}
\and S. Ali\thanksref{kansas}
\and N. M. Amin\thanksref{bartol}
\and K. Andeen\thanksref{marquette}
% ...
}
\authorrunning{IceCube Collaboration}
\thankstext{a}{also at Institute of Physics, Sachivalaya Marg, Sainik School Post, Bhubaneswar 751005, India}
\thankstext{b}{also at Department of Space, Earth and Environment, Chalmers University of Technology, 412 96 Gothenburg, Sweden}
\thankstext{c}{also at INFN Padova, I-35131 Padova, Italy}
\thankstext{d}{also at Earthquake Research Institute, University of Tokyo, Bunkyo, Tokyo 113-0032, Japan}
\thankstext{e}{now at INFN Padova, I-35131 Padova, Italy}
\institute{
% ...
Universit{\'e} Libre de Bruxelles, Science Faculty CP230, B-1050 Brussels, Belgium \label{brusselslibre}
% ...
\and Department of Physics, Loyola University Chicago, Chicago, IL 60660, USA \label{loyola}
\and Dept. of Physics and Astronomy, University of Canterbury, Private Bag 4800, Christchurch, New Zealand \label{christchurch}
% ...
\and Niels Bohr Institute, University of Copenhagen, DK-2100 Copenhagen, Denmark \label{copenhagen}
\and Dept. of Physics, TU Dortmund University, D-44221 Dortmund, Germany \label{dortmund}
% ...
\and Dept. of Physics and Astronomy, University of Kansas, Lawrence, KS 66045, USA \label{kansas}
% ...
\and Dept. of Physics and Wisconsin IceCube Particle Astrophysics Center, University of Wisconsin{\textemdash}Madison, Madison, WI 53706, USA \label{madisonpac}
% ...
\and Department of Physics, Marquette University, Milwaukee, WI 53201, USA \label{marquette}
% ...
\and Bartol Research Institute and Dept. of Physics and Astronomy, University of Delaware, Newark, DE 19716, USA \label{bartol}
% ...
\and Deutsches Elektronen-Synchrotron DESY, Platanenallee 6, D-15738 Zeuthen, Germany \label{zeuthen}
}
\date{Received: date / Accepted: date}
\maketitle
\twocolumn


\maketitle

\begin{abstract}
The IceCube Neutrino Observatory has observed a diffuse flux of high-energy astrophysical neutrinos for more than a decade. 
A relevant background to the astrophysical flux is prompt atmospheric neutrinos, originating from the decay of charmed mesons produced in cosmic-ray-induced air showers. 
The production rate of charmed mesons in the very forward phase space of hadronic interactions, and consequently, the prompt neutrino flux, remains uncertain and has not yet been observed by neutrino detectors. 
An accurate measurement of this flux would enhance our understanding of fundamental particle physics such as hadronic interactions in high-energy cosmic-ray-induced air showers and the nucleon structure. Furthermore, an experimental characterization of this background flux will improve the precision of astrophysical neutrino flux spectral measurements. 
In this work, we perform a combined fit of cascade-like and track-like neutrino events in IceCube to constrain the prompt atmospheric neutrino flux. 
Given that the prompt flux is a sub-dominant contribution, treating systematic uncertainties arising from the potential mis-modeling of the conventional and astrophysical neutrino fluxes is critical for its measurement. 
Our analysis yields a non-zero best-fit result, which is, however, consistent with the null hypothesis of no prompt flux within one standard deviation.
Consequently, we establish an upper bound on the flux at $4\times 10^{-16}$ (GeV m$^2$ s sr)$^{-1}$ at \SI{10}{TeV}.

\end{abstract}


\section{Introduction}

\begin{figure*}[t]
    \centering
    \includegraphics[width=0.8\linewidth]{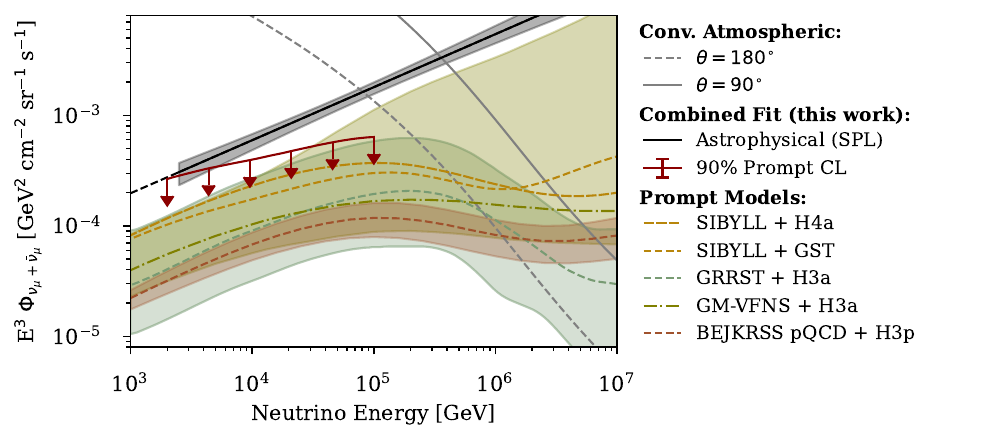}
    \caption{Model predictions for prompt neutrino fluxes. Shown are the models  SIBYLL \cite{Fedynitch:2018cbl}, GRRST \cite{Gauld:2015kvh}, GM-VFNS \cite{Benzke:2017yjn} and BEJKRSS \cite{Bhattacharya:2016jce}, as dash-dotted and dashed lines. Uncertainties for the latter three are plotted as colored bands in grey-green,  olive, and red, respectively. 
    The conventional neutrino flux \cite{Gaisser:2011klf,Fedynitch:2018cbl} (Conv.) from vertical zenith direction $\theta=180^\circ $  (dashed line) and horizontal zenith direction $\theta=90^\circ $  (solid line) direction is shown in grey.
    Also shown are the results of this analysis as an upper limit on the prompt flux (dark red) and the best-fit single power-law spectrum (SPL) of astrophysical neutrinos (black line).}
    \label{fig:PromptModels}
\end{figure*}

\begin{table}[bp]
    \centering
    \caption[Table of IceCube prompt flux results.]{Comparison of IceCube analyses of the prompt component. These analyses tested different prompt flux models, which mostly differ in normalization. The right column indicates the flux limit at \SI{10}{TeV}.}
    \label{tab:prompt_flux_results}
    \begin{tabular}{llll}
    \toprule
 Analysis & Prompt model & 90\% UL at \SI{10}{TeV} \\
    &  & [(GeV s cm$^2$ sr)$^{-1}$] \\
    \midrule
    \textit{6y Cascades} \cite{IceCube:2020acn} & BERSS \cite{Bhattacharya:2015jpa} & $3\cdot 10^{-16}$ \\
    \textit{6y NT} \cite{IceCube:2016umi}& ERS \cite{Enberg:2008te} & $3\cdot 10^{-16}$ \\
    \textit{7y HESE} \cite{IceCube:2020wum}& BERSS \cite{Bhattacharya:2015jpa} & $6\cdot 10^{-16}$\\
    \textit{ESTES} \cite{IceCube:2024fxo} & SIBYLL \cite{Fedynitch:2018cbl} & $6.4\cdot 10^{-16}$ \\
    \bottomrule
    \end{tabular}
\end{table}

Atmospheric neutrinos arise primarily from the decay of charged mesons produced in cosmic-ray-induced air showers. Below a few hundred GeV, the flux follows the primary cosmic-ray spectrum ($\propto E^{-2.7}$). Exceeding $\sim 1$ TeV, the increasing decay length of parent mesons leads to a steeper spectrum ($\propto E^{-3.7}$) as they re-interact before decaying.
Now, rarely produced charmed mesons ($D^{\pm}$, $D^0$) and baryons ($\Lambda_c$) become increasingly important sources of atmospheric neutrinos. 
 They decay ``promptly'' before reinteracting due to their short lifetimes. 
 The resulting prompt neutrino flux maintains the primary cosmic-ray spectral index up to PeV energies. 
 
 Figure \ref{fig:PromptModels} shows a comparison between the conventional, astrophysical, and prompt fluxes, outlining a central challenge: the prompt flux --- and the limit we are going to obtain --- is subdominant with respect to other components. For the conventional flux, the strong zenith dependence is illustrated by showing both the maximum (horizontal) and the minimum (vertical) flux. For the astrophysical flux, the best-fit single powerlaw from this analysis is shown, and for the prompt flux, the obtained upper limit and different model predictions are shown. 
 Towards lower energies, the conventional atmospheric flux dominates, making both prompt and astrophysical flux components challenging to measure. 
 Since the prompt flux follows the primary cosmic-ray spectral index, it is expected to overtake the conventional flux towards higher energies, but remains subdominant with respect to the astrophysical flux.

 To date, prompt atmospheric neutrinos have not been experimentally detected, and none of the preceding searches have observed a non-zero prompt flux \cite{IceCube:2020acn,IceCube:2016umi,IceCube:2020wum,IceCube:2024fxo,Abbasi:2021qfz,IceCube:2015mgt}. 
Table \ref{tab:prompt_flux_results} summarizes upper bounds from previous searches using different detection channels in IceCube. These bounds are of the order of the ERS prompt flux prediction \cite{Enberg:2008te}.

 For the flux expectation, significant uncertainties exist in the charm production cross-sections \cite{Jeong:2017pac}, particularly in the forward region where particles are produced at small angles relative to the cosmic-ray direction.
Models such as SIBYLL 2.3c \cite{Fedynitch:2018cbl}, GRRST \cite{Gauld:2015kvh},  GMS \cite{Benzke:2017yjn,Garzelli:2015psa}, BERSS/BEJKRSS \cite{Bhattacharya:2016jce,Bhattacharya:2015jpa}, and the aforementioned ERS model \cite{Enberg:2008te} predict an inclusive cross-section for the relevant mesons. The leading-order production mechanisms are gluon-gluon fusion producing a $c\bar{c}$ pair and interactions between a gluon and a sea charm-quark.
This can be combined with a model of the primary nucleon flux using a semi-analytical solution of the cascade equations \cite{Gaisser2016} or a numerical solver like MCEq \cite{Fedynitch:2015zma}. Figure \ref{fig:PromptModels} shows that the models vary by almost a factor of 10 but exhibit rather similar spectral shapes, following the primary cosmic-ray nucleon spectrum.

The prompt atmospheric neutrino flux is expected to be largely isotropic, and its energy spectrum is roughly one power harder than that of conventional atmospheric neutrinos.
Because it is more similar to astrophysical neutrinos than other backgrounds, the unknown prompt flux introduces considerable uncertainty in any measurements of the astrophysical neutrino spectrum and vice versa; conversely, features of the astrophysical spectrum can be falsely attributed to a contribution from prompt neutrinos.

The sensitivity to the prompt atmospheric neutrino flux is significantly improved via a combined analysis using numerous IceCube detection channels and enlarged exposure. 
This study combines horizontal and up-going tracks (Northern Tracks) \cite{Abbasi:2021qfz} and cascades (Cascades) \cite{IceCube:2020acn} using the Combined Fit method as outlined in \cite{IceCube:2025tgp}. The IceCube Neutrino Observatory and these event selections are introduced in Section \ref{sec:DetectorSample}. 
We outline the Combined Fit and our study of the prompt atmospheric neutrino flux in Section \ref{sec:analysismethod}. Section \ref{sec:sensitivity} presents the sensitivity of the Combined Fit and systematic tests regarding the background components, followed by the results of the prompt neutrino flux analysis in Section \ref{sec:Results} and an outlook in Section \ref{sec:Conclusion}.

\section{Track and Cascade Selections in IceCube}\label{sec:DetectorSample}

The IceCube Neutrino Observatory detects neutrinos through Cherenkov light emission from secondary particles produced in neutrino interactions. The cubic-kilometer detector consists of 5160 digital optical modules (DOMs) arranged on 86 strings embedded in the Antarctic ice \cite{IceCube:2016zyt}. 

To optimize sensitivity across different neutrino interaction types, IceCube employs multiple complementary event selections. The Northern Tracks selection \cite{Abbasi:2021qfz,IceCube:2016umi} focuses on up-going and horizontal neutrino-induced muon tracks, providing excellent angular resolution. 
By focusing on up-going and horizontal zenith directions ($\theta > 85^{\circ}$), cosmic-ray muons are absorbed in the Earth and the Antarctic ice sheet before reaching the detector. The effective detection volume for muon neutrinos is significantly greater than the geometric detector volume because high-energy muons penetrate through the ice for several kilometers. 
The energy measurement is based on the energy loss of muons inside the detector volume \cite{IceCube:2012iea}. Due to the unknown distance the muon has traveled before entering, it is only a lower bound on the actual neutrino energy.

The cascade selection \cite{IceCube:2020acn} targets neutrino interactions inducing a hadronic or electromagnetic cascade inside the detector volume. 
Unlike muon tracks, cascades develop within a small volume with a typical elongation of \SIrange{10}{20}{m} along the direction of the neutrino. Cherenkov light is emitted omnidirectionally, with higher intensity in the forward direction.
For cascade-like events from charged-current (CC) neutrino interactions, the hadronic cascade and the outgoing lepton are superimposed, so the full neutrino energy is deposited. 
For neutral current events (NC), a fraction of the neutrino's energy is carried away by the outgoing neutrino. Atmospheric muons enter the detector from above and the side. 
In the cascade selection, this background is rejected by vetoing events entering the detector from outside. This veto also leads to the rejection of any atmospheric neutrino accompanied by synchronously detected atmospheric muons, an effect which is called self-veto \cite{Gaisser:2014bja,Arguelles:2018awr}. 
This leads to an enhanced purity of astrophysical neutrinos above TeV energies in the down-going region.

Both selections apply additional cuts based on machine learning tools (Boosted Decision Trees) to further reduce the contamination from misreconstructed atmospheric muons. Both selection channels are well established in previous analyses measuring the spectrum of astrophysical neutrinos \cite{IceCube:2020acn,Abbasi:2021qfz,IceCube:2025tgp,IceCube:2016umi,IceCube:2025ewu}. The complementary strengths of these selections - tracks: high statistics and 
 cascades: good energy resolution, the complementary angular acceptance, and different backgrounds motivate a combined analysis as employed here. This combined approach is detailed in \cite{IceCube:2025tgp,IceCube:2025ewu}. It is also expected to enhance our ability to disentangle the prompt atmospheric component from the other flux contributions.

\begin{figure}[htp]
    \centering
    \includegraphics[width=\linewidth]{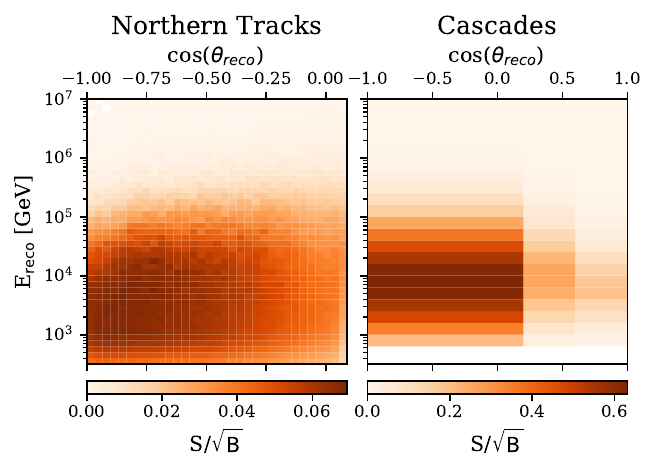} 
    \caption{Ratio of prompt neutrino rate $S$ over the square root of the total background rate $B$ from conventional atmospheric and astrophysical neutrinos and  atmospheric muons in the track and 
    cascade samples. The axes correspond to the reconstructed zenith $\theta$ and energy $E$ of the respective selection. The choice of bins is identical to \cite{IceCube:2025tgp,IceCube:2025ewu}.}
    \label{fig:RecoSignificance}
\end{figure}

\section{Analysis Method}
\label{sec:analysismethod}

The analysis method of the Combined Fit is based on forward folding and utilizes a binned effective likelihood $\mathcal{L}$ \cite{Arguelles:2019izp} comparing templated model expectations with the observed experimental counts. 
For this, we bin the data for both detection channels in two-dimensional histograms of the reconstructed energies and zenith directions as displayed in Figure
 \ref{fig:RecoSignificance}. 
The expectations in each bin consist of the sum of weighted Monte-Carlo (MC) neutrino event simulations. Each MC event weight depends on the sum of all flux components evaluated at its true energy and direction. 
The relevant components which are expected to contribute to the data samples above \SI{100}{GeV} are: i) conventional atmospheric neutrinos, ii)  astrophysical neutrinos, iii) prompt atmospheric neutrinos, and iv)  atmospheric muons.
Each of these components is parameterized based on model assumptions and scaled by a normalization parameter $\phi^0_{\mathrm{component}}$. We then maximize the likelihood $\mathcal{L}$ for all our parameters (introduced below) and do profile likelihood scans to estimate the confidence intervals of our signal parameters.
In general, the fraction of prompt atmospheric neutrinos does not exceed 10\% of the total neutrino rate anywhere in the phase space of the analysis bins. Therefore, careful treatment is required for modeling the dominant background components, as mismodeling can lead to biases in our signal estimation (see \cite{IceCube:2023mrq}).

Both of the atmospheric neutrino components, conventional and prompt,  are based on estimations from the cascade equation solver MCEq using the SIBYLL 2.3c \cite{Fedynitch:2018cbl} ha\-dro\-nic interaction model and the Gaisser H4a \cite{Gaisser:2011klf} cosmic-ray flux model. The hadronic model uncertainties follow the para\-me\-tri\-zation made by Barr et al.\  \cite{Barr:2006it}, by using the most relevant parameters ($h, w, x, y$). This description also covers uncertainties in the $\nu/\bar{\nu}$-flux ratio \cite{Yanez:2023lsy}. Primary cosmic-ray model uncertainties are parameterized by a variation of the spectral index $E^{-\Delta\gamma_{CR}}$ and a linear interpolation between the H4a \cite{Gaisser:2011klf} and GST \cite{Gaisser:2013bla} primary flux models: $\Lambda_{\mathrm{CR}} \phi_{\mathrm{GST}}+ (1-\Lambda_{\mathrm{CR}})\phi_{\mathrm{H4a}}$. 
This interpolation is consistently applied to both the conventional and prompt predictions.

To account for the astrophysical neutrino flux uncertainties, we employ five different parameterizations:
\begin{enumerate}
    \item Single power-law (SPL): \newline
    $\phi_{\nu}=\phi_{\mathrm{astro}}^0 \left(\frac{E_{\nu}}{\mathrm{100 TeV}}\right)^{-\gamma_{\mathrm{astro}}}$

    \item SPL + cutoff: \newline 
    $\phi_{\nu} = \phi_{\mathrm{astro}}^0 \left(\frac{E_{\nu}}{\mathrm{100 TeV}}\right)^{-\gamma_{\mathrm{astro}}}\cdot e^{-E_{\nu}/E_{\mathrm{cutoff}}}$

    \item Broken power-law (BPL): \newline 
    Powerlaw flux, with $\gamma_{\mathrm{astro}}=\gamma_1$ for  $E < E_{\mathrm{break}}$ and $\gamma_{\mathrm{astro}} =\gamma_2$ for  $E > E_{\mathrm{break}}$

    \item Log-parabola (Log-P):\newline 
    $\phi_{\nu} = \phi_{\mathrm{astro}}^0 \left(\frac{E_{\nu}}{\mathrm{100 TeV}}\right)^{-\alpha - \beta (E_{\nu}/100 \mathrm{TeV})}$

    \item Segmented Powerlaw: \newline 
    14 $E^{-2}$ pieces, with each having an individual normalization. Each order of magnitude in energy is divided into three segments, except for the first (\SI{100}{GeV} - \SI{1}{TeV}), which has two segments, and the last (\SI{10}{PeV} - \SI{100}{PeV}), which has one segment.
\end{enumerate}
Detailed descriptions and the corresponding fit results of these models can be found in \cite{IceCube:2025ewu}. 

The detector uncertainties are implemented based on bin-wise gradients calculated from random variation of each detector uncertainty parameter \cite{IceCube:2019lxi}. The uncertainties include:
\begin{itemize}
    \item Bulk ice scattering and absorption length ($\epsilon_{\mathrm{scatt}}, \epsilon_{\mathrm{abs}}$) \cite{IceCube:2013llx}
    \item Optical efficiency of the modules ($\epsilon_{\mathrm{DOM}}$) \cite{IceCube:2016zyt}
    \item Acceptance of the borehole ice ($p_0$ and $p_1$) \cite{IceCube:2023ahv}
\end{itemize}

The remaining cosmic-ray muon background is handled by scaling the templates for each selection with a respective normalization parameter $\phi_{\mu;NT}^0 $ and $\phi_{\mu;Casc}^0 $ \cite{IceCube:2025ewu}. Here, we also account for the uncertainties associated with the energy threshold $\epsilon_{veto} $ of the atmospheric neutrino self-veto in the down-going region for the cascade selection. Neutrino cross-sections, including inelasticities, are based on the CSMS model \cite{Cooper-Sarkar:2011jtt}.

We investigate the 
sensitive region of bins contributing to the overall sensitivity of the analysis by comparing the signal expectation with the different background components. 
Figure \ref{fig:RecoSignificance} shows the ratio of expected prompt signal rate to the expected square root of the background rate for each analysis bin of the two selections. The largest ratio is located in the up-going zenith region. In reconstructed energy, the highest ratio is between a few TeV and tens of TeV. In no region of the phase space is the prompt flux dominant.

\section{Prompt Flux Sensitivity and Systematic Studies}\label{sec:sensitivity}

The prompt atmospheric neutrino flux is parametrized by the baseline expectation from the Sybill model multiplied with a free normalization factor $\phi_{\mathrm{prompt}}(E) = \phi^0_{\mathrm{prompt}}\cdot \phi_{\mathrm{SIBYLL}}(E)$.
To evaluate the sensitivity to the flux, we construct a confidence interval by calculating the log-likelihood differences between test values of the prompt flux normalization $\phi^0_{\mathrm{prompt}}$ and the best-fit value of the normalization $\hat{\phi}^0_{\mathrm{prompt}}$. Here, we follow the procedure of  \cite{Cowan:2010js} and evaluate the significance with a so-called Asimov fit 
of the injected null hypothesis of no prompt flux: 
$\phi^0_{\mathrm{prompt}}=0$ and baseline values for the other parameters, including the best-fit SPL values from \cite{IceCube:2025tgp} for the astrophysical flux. We convert log-likelihood differences to p-values assuming that $-2\Delta \log \mathcal{L}$ is $\chi^2$ distributed. For a bounded parameter such as the normalization, this assumption must be corrected as discussed in \cite{Cowan:2010js} in order to construct Feldman-Cousin \cite{Feldman:1997qc} based confidence intervals. 

\begin{figure}[htpb]
    \centering
    \includegraphics[width=\linewidth]{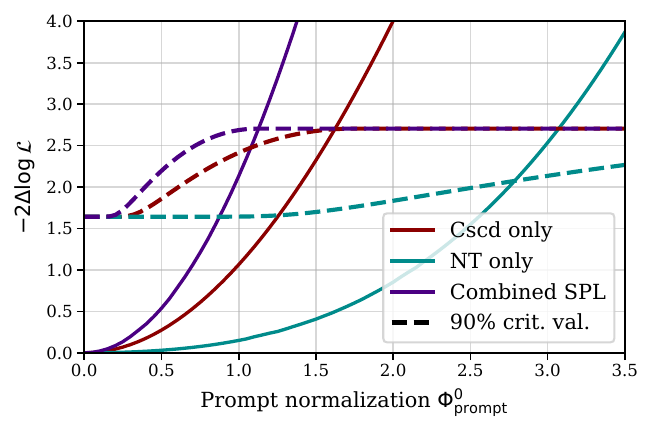}
    \caption{Likelihood difference of the best-fit to the injected null hypothesis $\phi_{\mathrm{prompt}} = 0$ and tested non-zero prompt flux normalizations. This is shown for the individual samples (cascades in red, Northern Track in turquoise) and the Combined Fit in blue. The dashed line corresponds to the critical value of 90\% confidence after correction (see text). Sensitivities correspond to the crossing point of the likelihood profile and the critical value line.}
    \label{fig:Sample_contours}
\end{figure}

Assuming the SPL astrophysical neutrino flux,
we show in Figure \ref{fig:Sample_contours} the resulting sensitivities for the combined samples and for each of the two selections individually. The Cascade selection is more sensitive than the Northern Track selection, and combining these selections further improves the sensitivity.

\begin{figure}[htpb]
    \centering
    \includegraphics[width=\linewidth]{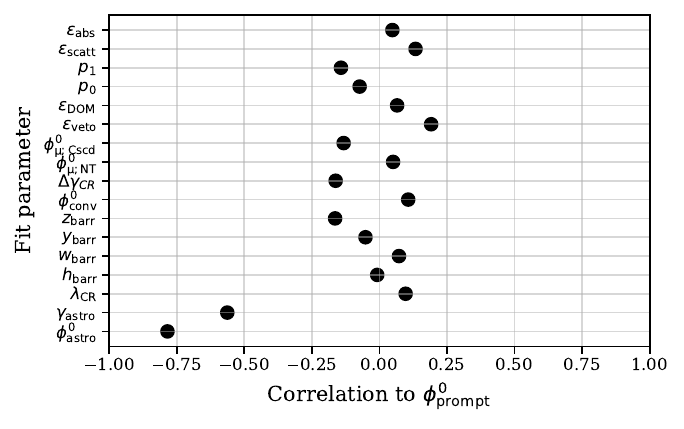}
    \caption{Correlation coefficient of the prompt flux normalization with the other fit parameters. 
    }
    \label{fig:prompt_corr}
\end{figure}

Independent of the specific model assumption, we expect that the prompt flux contributes only a small fraction to the total neutrino rate in our analysis. Therefore, it is important to understand the correlations of the fitted normalization with other fit parameters of the analysis. For this we perform an Asimov fit similar to the above, but with the prompt normalization set to $\phi^0_{\mathrm{prompt}}=1$. Figure \ref{fig:prompt_corr} shows that the prompt normalization is correlated with the parameters of the astrophysical flux but is largely uncorrelated with other parameters.

\begin{figure}[tpb]
    \centering
    \includegraphics[width=0.7\linewidth]{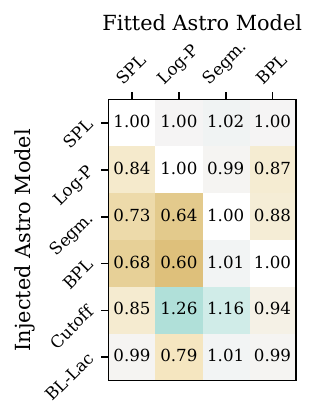}
    \caption{ Test of the bias on $\phi^0_{\mathrm{prompt}}$ introduced by the assumed astrophysical model parametrization in the fit. The columns correspond to the fitted astrophysical model; the rows to the injected model. A value closer to the injected prompt normalization $\phi^0_{\mathrm{prompt}}=1$, corresponds to a smaller bias caused by fitting the astrophysical model of the respective column. Model definitions are in the text; for the cutoff and BL-Lac models, please refer also to \cite{Abbasi:2021qfz} and \cite{Padovani:2015mba}.}
    \label{fig:Astro_confusion}
\end{figure}

As a critical aspect of this analysis, we have to investigate how a specific assumed parameterization of the astrophysical flux results in a biased prompt normalization. We already have evidence from previous studies \cite{IceCube:2025tgp} that the astrophysical flux has a spectral shape beyond the SPL.
We test this bias by injecting and fitting different models of the astrophysical flux and evaluating the deviation of the best-fit prompt flux normalization from the injected value of $\phi_{\mathrm{prompt}}^0=1$. Figure \ref{fig:Astro_confusion} shows the resulting prompt normalizations of this test.
We observe substantial biases in the prompt normalization if the injected spectrum deviates from the fitted astrophysical flux model. Independent of the injected model, the smallest bias is obtained for the segmented piece model, as expected from the large degree of freedom of the model. Therefore, we use the segmented piece model as our reference astrophysical flux model in the analysis.

In addition to the parameterization of the spectral shape of the astrophysical flux, we test how the known anisotropic neutrino flux from the galactic plane \cite{IceCube:2023ame} impacts the prompt flux measurement. 
In our tests, we have injected the best fits \cite{IceCube:2023hou} for either the normalization of the Fermi-$\pi^0$ model \cite{Fermi-LAT:2012edv} or the  CRINGE flux model \cite{Schwefer:2022zly}, and then we performed an Asimov fit without including such a galactic component. Both tests result in a reduction of the fitted prompt flux normalization by 20\%. In the experimental analysis, a galactic plane fit is included as an additional test, as discussed below.

\begin{table}[htpb]
    \centering
    \caption{Fitted prompt flux normalizations for different injected atmospheric neutrino flux models. The injected prompt normalization is $\phi^0_{\mathrm{prompt}}=1$. The injected conventional flux is given in each row (Gaisser H4a \cite{Gaisser:2011klf}, GST \cite{Gaisser:2013bla}, GSF \cite{Dembinski:2017zsh}, Honda-2006 \cite{Honda:2006qj}, \textsc{Daemonflux} \cite{Yanez:2023lsy}). We fit with the baseline uncertainty treatment based on the Barr-scheme \cite{Barr:2006it} and the H4a-GST interpolation (left column) and the \textsc{Daemonflux} treatment (right column), see text.}
    \label{tab:atmo_bias}
    \begin{tabular}{lrr}
    \toprule
 Injected Model & H4a-GST & \textsc{Daemonflux}  \\
    \midrule
 H4a & 1.00 & 0.84 \\
 GST & 0.98 & 0.10 \\
 GSF & 1.11 & 0.87 \\
 Honda & 0.70 & 0.97 \\
 \textsc{Daemonflux} & 0.73 & 1.00 \\
    \bottomrule
    \end{tabular}
\end{table}

\begin{figure*}[htpb]
    \centering
    \includegraphics[width=\linewidth]{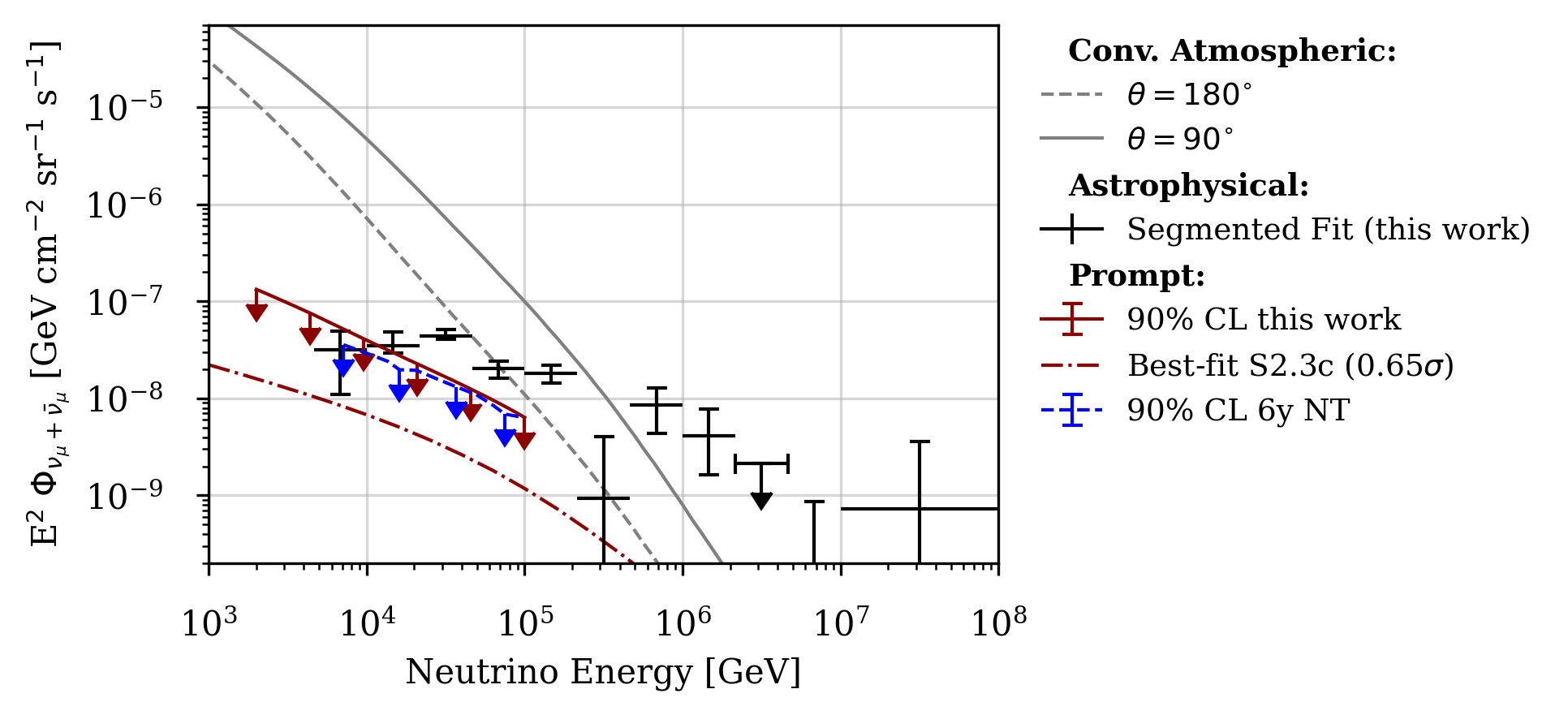}
    \caption{The limit on the prompt atmospheric neutrino flux (in dark red, cf. also Figure \ref{fig:PromptModels}) compared to the previous 6-year NT-limit  (blue dashed line) \cite{IceCube:2016umi}. The limits are given in their respective sensitive energy ranges. The best-fit prompt flux is also shown, the fitted segmented astrophysical flux in black, and the conventional atmospheric neutrino flux as a gray line.}
    \label{fig:Upper_Lim}
\end{figure*}

Similar to the above bias tests, we test whether our parameterization of atmospheric flux uncertainties can impact the prompt measurement. This time, we inject several different atmospheric flux assumptions with varying primary flux models and test if our chosen parameterization of hadronic and primary cosmic-ray flux uncertainties covers these uncertainties without biasing the prompt flux normalization.
As an additional comparison, we switch our baseline fit model 
to the data-driven \textsc{Daemonflux} model \cite{Yanez:2023lsy}. This entails a replacement of the model assumptions, namely SIBYLL 2.3c and Gaisser H4a, together with the associated uncertainty treatment (Barr parameters, $\Delta\gamma$, and $\lambda_{CR}$), by the parameterizations presented in \cite{Yanez:2023lsy}. Table \ref{tab:atmo_bias} shows the result of these tests. Generally, our baseline fit model H4a-GST is robust with a bias less than 30\% on the fitted prompt normalization for different injected conventional fluxes.
This picture changes when fitting with the recent DEAMONFlux model, which shows a stronger bias.

In addition to the above tests, we have performed fits with additional systematic parameters describing effects not included in the baseline fit. 
Here, we follow the procedure and parameter definitions from \cite{IceCube:2025ewu} and include a flux normalization parameter for neutrinos from the galactic plane, a parameter scaling the total neutrino interaction cross-section, a parameter scaling the muon energy loss, and a parameter scaling the inelasticity of the neutrino interaction.

\section{Results}\label{sec:Results}

\begin{table}[b]
    \centering
    \caption[Table of the atmospheric model bias on the prompt flux normalization.]{Fit results for the prompt flux for different astrophysical flux models and for further tests. The baseline assumption for these is a conventional atmospheric model based on the SIBYLL 2.3c hadronic interaction model and the Gaisser H4a primary cosmic-ray model (except for the \textsc{Daemonflux} test). The test with additional systematics (Add. sys.) includes the additional systematic parameters mentioned in the text. The best fit and the confidence intervals are prompt flux normalizations. Confidence intervals are constructed by the Feldman-Cousins method \cite{Feldman:1997qc} and assume the bounded $\chi^2$ distribution in \cite{Cowan:2010js}. The significance (sign.) is based on the $-2\Delta \log\mathcal{L}$ of the best fit towards the null hypothesis and converted into sigma assuming it follows a $\chi^2$ distribution \cite{Wilks:1938dza}. }\label{tab:astro_sys_tests}
    \begin{tabular}{lrrrr}
    \toprule
 Model Assumption & best-fit & 90\% CL & 1$\sigma$ interval & sign. [$\sigma$] \\
    \midrule
 Segm. & 0.76 & 2.59 & (0.11, 1.86) & 0.65 \\
 SPL & 0.23 & 1.92 & (0.00, 1.25) & 0.22 \\
 Log-P & 0.22 & 2.00 & (0.00, 1.31) & 0.18 \\
 BPL & 1.15 & 2.99 & (0.26, 2.29) & 0.94 \\
 \hline
 Add. sys. & 1.07 & 2.83 & (0.24, 2.13) & 0.94 \\
 Cascades only & 0.00 & 1.69 & (0.00, 0.74) & 0.00 \\
 NT only & 0.00 & 2.38 & (0.00, 0.70) & 0.00 \\
 \textsc{Daemonflux} & 1.17 & 3.00 & (0.27, 2.53) & 0.87 \\
    \bottomrule
     \end{tabular}
\end{table}

The results of the fit of the experimental data are listed in Table \ref{tab:astro_sys_tests}, with best fits and confidence intervals for each fitted model assumption. The baseline assumption gives a non-zero best fit of the prompt normalization. However, with a significance of only $0.65\sigma$, this is consistent with the null hypothesis of no prompt flux. 
The corresponding 90\% upper limit is $\phi_{\mathrm{prompt}}^0 = 2.59$ and is shown in Figure \ref{fig:Upper_Lim}  together with the astrophysical and conventional atmospheric components. Variations of the astrophysical models are consistent with the baseline result within the uncertainties. Additionally, the table shows the results of the further tests as described above. The test of the added systematics (Add. sys. in the table) increases the fitted prompt normalization despite an increased number of parameters. The likelihood of this test is, however, not significantly improved.

Changing the atmospheric flux model to \textsc{Daemonflux}  also increases the result of the prompt normalization. Note that \textsc{Daemonflux}  also changes the prompt prediction due to the change of the primary cosmic-ray model which makes a comparison difficult. However, the significance remains well below $1\sigma$ and is comparable with our baseline result.

Fitting the prompt flux normalization with only the individual samples results in a best fit of $\phi^0_{\mathrm{prompt}}=0$ in both cases. This result is consistent with previous observations \cite{IceCube:2020acn,IceCube:2016umi}. Therefore, the emergence of a non-zero best fit by combining the samples is not intuitively expected. The effect can be understood as a benefit of the combination of data samples with consistent treatment of systematic uncertainties: the high-statistics Northern Tracks sample largely constrains the conventional flux model, and this also changes the conventional flux assumption in the down-going region of the cascade sample. This, in turn, leads to an increased and better constrained self-veto threshold, which then allows for a larger flux of prompt atmospheric neutrinos. The normalization of the astrophysical flux in the fit of the combined sample is unchanged with respect to \cite{IceCube:2025tgp,IceCube:2025ewu} and shows no significant decrease with respect to the fit of the individual samples. The insignificant observation of a non-zero flux of prompt neutrinos does not affect the highly significant observation of the astrophysical flux.

\section{Conclusion and Outlook}\label{sec:Conclusion}

We present an updated search for prompt atmospheric neutrinos. With respect to previous analyses, we combine two data samples, IceCube's Cascade and Northern Tracks, with consistent treatment of systematic uncertainties.
This Combined Fit improves the sensitivity to a prompt flux and, more importantly, enhances the robustness with respect to the unknown astrophysical flux spectrum and other uncertainties compared to previous results. The best fit results in a non-zero but statistically insignificant flux normalization. Figure \ref{fig:Upper_Lim} shows the corresponding 90\% upper limit and best fit of the prompt flux; see also Figure \ref{fig:PromptModels}.
 The upper limit starts constraining the theoretical predictions, but does not rule out current models. The best fit is consistent with the baseline SIBYLL 2.3c model.

The resulting limit coincides with the previously reported limit in \cite{IceCube:2016umi}. However, that previous result was based on an assumed power-law astrophysical flux model, which we have replaced with the more flexible segmented powerlaw model. 
Furthermore, the previous result was based on a fitted under-fluctuation, i.e.\ a vanishing prompt flux for the best fit, while we observe a non-zero best fit in this analysis, which corresponds to a statistically weaker upper bound.

A measurement of the prompt neutrino flux component in the diffuse spectrum of neutrinos will remain challenging, as the analysis is limited by the uncertainties of the overwhelming background components. The unknown spectrum of the astrophysical neutrino flux will likely pose a challenge also to future searches for prompt neutrinos. However, the strategy of combining complementary data selections is established as a promising strategy also for future work. Several selections can be considered, such as, e.g.,\ ESTES \cite{IceCube:2024fxo}, MESE \cite{IceCube:2025tgp}, and DNN-Cascades \cite{IceCube:2023ame}, in addition to an extension of the lifetime of the here-used selections.

In addition to the aforementioned combinations, one can envision a combination with high-energy atmospheric muon measurements.

Improved modeling of the conventional neutrino flux by collider measurements of meson production, improved primary cosmic ray measurements, and improved constraints from atmospheric muon measurements, as implemented in \textsc{Daemonflux}  \cite{Yanez:2023lsy}, would be highly beneficial for better constraint backgrounds.

\begin{acknowledgements}
The IceCube Collaboration acknowledges the significant contributions to this manuscript from Jakob Böttcher, Philipp Fürst, and Christopher Wiebusch. The authors gratefully acknowledge the support from the following agencies and institutions:
USA {\textendash} U.S. National Science Foundation-Office of Polar Programs,
U.S. National Science Foundation-Physics Division,
U.S. National Science Foundation-EPSCoR,
U.S. National Science Foundation-Office of Advanced Cyberinfrastructure,
Wisconsin Alumni Research Foundation,
Center for High Throughput Computing (CHTC) at the University of Wisconsin{\textendash}Madison,
Open Science Grid (OSG),
Partnership to Advance Throughput Computing (PATh),
Advanced Cyberinfrastructure Coordination Ecosystem: Services {\&} Support (ACCESS),
Frontera and Ranch computing project at the Texas Advanced Computing Center,
U.S. Department of Energy-National Energy Research Scientific Computing Center,
Particle astrophysics research computing center at the University of Maryland,
Institute for Cyber-Enabled Research at Michigan State University,
Astroparticle physics computational facility at Marquette University,
NVIDIA Corporation,
and Google Cloud Platform;
Belgium {\textendash} Funds for Scientific Research (FRS-FNRS and FWO),
FWO Odysseus and Big Science programmes,
and Belgian Federal Science Policy Office (Belspo);
Germany {\textendash} Bundesministerium f{\"u}r Bildung und Forschung (BMBF),
Deutsche Forschungsgemeinschaft (DFG),
Helmholtz Alliance for Astroparticle Physics (HAP),
Initiative and Networking Fund of the Helmholtz Association,
Deutsches Elektronen Synchrotron (DESY),
and High Performance Computing cluster of the RWTH Aachen;
Sweden {\textendash} Swedish Research Council,
Swedish Polar Research Secretariat,
Swedish National Infrastructure for Computing (SNIC),
and Knut and Alice Wallenberg Foundation;
European Union {\textendash} EGI Advanced Computing for research;
Australia {\textendash} Australian Research Council;
Canada {\textendash} Natural Sciences and Engineering Research Council of Canada,
Calcul Qu{\'e}bec, Compute Ontario, Canada Foundation for Innovation, WestGrid, and Digital Research Alliance of Canada;
Denmark {\textendash} Villum Fonden, Carlsberg Foundation, and European Commission;
New Zealand {\textendash} Marsden Fund;
Japan {\textendash} Japan Society for Promotion of Science (JSPS)
and Institute for Global Prominent Research (IGPR) of Chiba University;
Korea {\textendash} National Research Foundation of Korea (NRF);
Switzerland {\textendash} Swiss National Science Foundation (SNSF).
\end{acknowledgements}

\bibliographystyle{spphys}       
\bibliography{references}   

\end{document}